# High-repetition-rate source delivering optical pulse trains with a controllable level of amplitude and temporal jitters

Ugo Andral [1] and Christophe Finot *[1]

Laboratoire Interdisciplinaire CARNOT de Bourgogne, UMR 6303 CNRS-Université de Bourgogne-Franche Comté

christophe.finot@u-bourgogne.fr

We theoretically propose and numerically validate an all-optical scheme to generate optical pulse trains with varying peak-powers and durations. A shaping of the spectral phase thanks to discrete π/2 phase shifts enables an efficient phase-to-intensity conversion of a temporal phase modulation based on a two-tone sinusoidal beating. Experiments carried out at telecommunication wavelengths and at a repetition rate of 10 GHz confirm the ability of our approach to efficiently generate a train made of pulses with properties that vary from pulse-to-pulse. The levels of jitters can be accurately controlled.

## 1  |  INTRODUCTION

The generation of pulse trains at repetition rates of several tens of GHz remains a crucial step for the testing of optical components. In order to assess the performance of devices under realistic conditions, it is relevant to have a source which quality can be adjusted to mimic the presence of various kinds of degradations that may impair the optical function under test. Whereas the optical signal-to-noise ratio can be straightforwardly adjusted using an external source of amplified spontaneous emission, jitters of the peak-power and duration of the pulses are more challenging to control, especially when these fluctuations are on a short time scale and affect pulses with a low duty cycle. Indeed, the current bandwidth limitations of optoelectronic devices do not allow the direct generation of picosecond pulses directly from intensity modulators. As a consequence, alternate solutions are required to generate the pulse train such as actively mode-locked fiber lasers [1] or cavity-free techniques based on the nonlinear reshaping of a sinusoidal beating [2, 3]. These approaches have in common a very high stability and the resulting pulses are perfectly defined without any fluctuations of their temporal profile, which may limit their use for component testing applications. Another attractive solution is based on a direct temporal phase modulation that is then converted into an intensity modulation thanks to a quadratic spectral phase [4, 5]. Picosecond pulses at repetition rates of several tens of GHz have been successfully demonstrated and such a source can be involved in optical sampling [6]. However, this approach suffers from a limited extinction ratio and from the presence of detrimental temporal sidelobes [7], so that a considerable part of the energy lies outside the main pulse. By replacing the quadratic spectral phase modulation by a triangular one, we have recently shown that it was possible to achieve the generation Fourier-transform limited, close-to-Gaussian pulses [8] that are well suited for return-to-zero transmissions.

We extend here theoretically and experimentally this scheme and we demonstrate that an architecture based on a dual tone sinusoidal phase modulation can efficiently sustain the generation of a train of pulses with peak-powers and durations that vary from pulse-to-pulse. The rate of change as well as the level of fluctuations can be easily controlled by adjusting the RF properties of the second tone.

Our article is therefore organized as follows. We first describe the principle of our approach before assessing by numerical simulations and by an approximate analytical model the ultrafast fluctuations that can be imprinted on the pulse train. We then validate our optical architecture experimentally for a pulse train at a repetition rate of 10 GHz.

## 2 | PRINCIPLE OF OPERATION AND NUMERICAL SIMULATIONS

### 2.1 | PULSE GENERATION PROCESS

Before explaining our approach, let us first recall the basis of the generation of a stable pulse train from a temporal sinusoidal phase modulation as recently introduced in [8]. We consider a continuous optical wave with a power $P_0$ and an optical carrier frequency $\omega_c$, $\Psi(t) = \sqrt{P_0}\, \psi(t)\, e^{i\,\omega_c\, t}$, which phase is temporally modulated by a sinusoidal waveform $\varphi(t)$:

$$\begin{cases} \psi(t) = e^{i\,\varphi(t)} \\ \varphi(t) = A_m \cos(\omega_m t) \end{cases} \tag{1}$$

where $A_m$ is the amplitude of the phase modulation and $\omega_m$ its angular frequency. The temporal sinusoidal phase leads to a set of spectral lines that are equally spaced by $\omega_m$ and which amplitude can be expressed using a Jacobi-Anger expansion [9, 10]:

$$\psi(t) = \sum_{n=-\infty}^{\infty} i^n\, J_n(A_m)\, e^{i\,n\,\omega_m\,t}, \tag{2}$$

with $J_n$ being the Bessel function of the first kind of order $n$. An essential point is the existence of a phase shift of $\pi/2$ between each spectral component. With the progress of linear shaping, one can now process a line-by-line spectral phase profile using liquid-crystal modulators [11] or fiber Bragg gratings [12]. It then becomes feasible to imprint an exact spectral phase profile of opposite sign. Quite remarkably, this spectral phase profile to be applied in order to obtain a flat spectral phase does not depend on the value of $A_m$. The resulting temporal profile is a Fourier-transform limited waveform $\psi'(t)$ at a repetition rate $f_m = \omega_m/2\pi$ (leading to a period $T_0 = 1/f_m$) that can be expressed as:

$$\psi'(t) = J_0(A_m) + 2\sum_{n=1}^{\infty} J_n(A_m)\, \cos(n\,\omega_m\,t). \tag{3}$$

We have reported in Fig. 1(a) the resulting intensity profiles that are obtained for various values of $A_m$. According to the amplitude of the initial temporal phase modulation $A_m$, the temporal duration, the peak-power and the overall pulse shape vary significantly. However, as the phase-to-intensity conversion process is not dissipative, let us note that all the waveforms have the same energy over a period $T_0$. The value $A_m = 1.1$ rad (Fig. 1a2) is of high interest as a clean Gaussian-like structure is achieved, with a high extinction ratio [8]. For lower values such as $A_m = 0.6$ rad (Fig. 1a1), a residual background is visible and the shape is close to an Akhmediev breather [13]. For values of $A_m$ higher than 1.1 rad (Fig. 1a3), the extinction ratio is decreased but shorter durations and higher peak-power can be reached. The detailed evolution of the peak-power and full-width at half maximum (fwhm) duration is provided in Fig. 1(b-c). We can note that these quantities evolve monotonously and that they can vary over a factor above 4. Around $A_m = 1.1$ rad, it is possible to approximate their evolution by linear fits for the peak-power and the duration respectively (dashed grey lines).

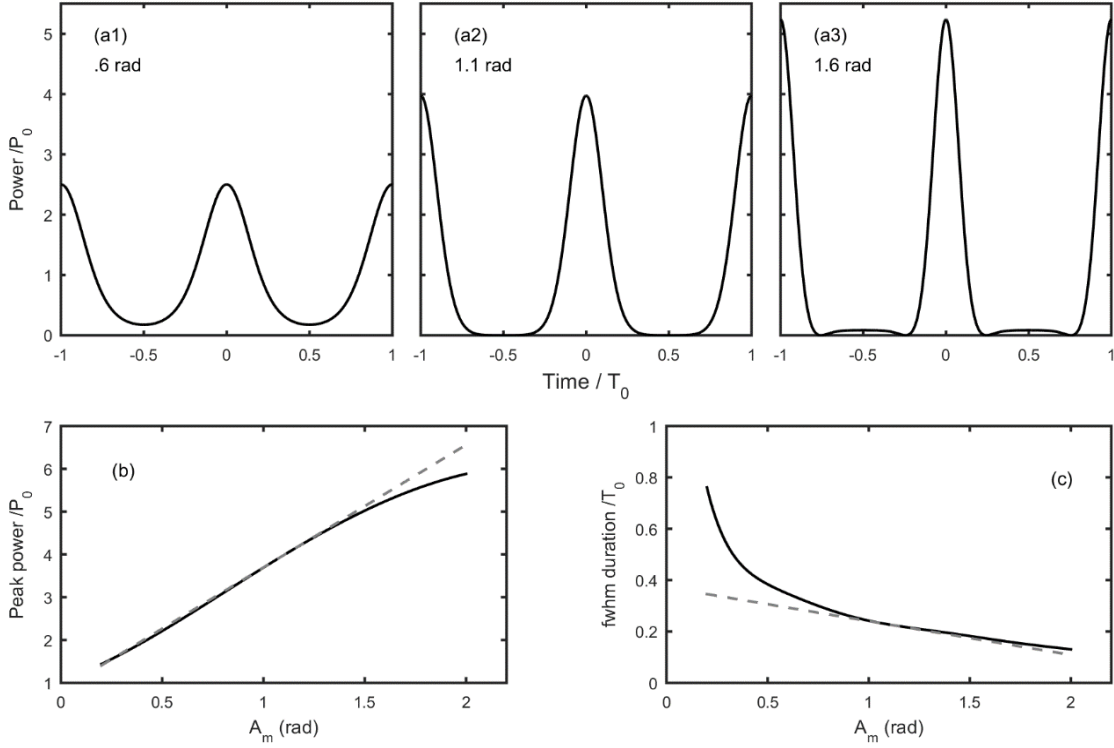

**Figure 1** (a) Temporal intensity profiles obtained for different amplitude of phase modulation ($A_m = .6$, 1.1 and 1.6 rad, panels 1, 2 and 3 respectively). Evolution of the peak-power and the temporal duration of the resulting structure (panels b and c, respectively). Results of numerical simulations (solid black lines) are compared with a Taylor expansion around $A_m = 1.1$ rad (dashed grey line).

## 2.2 | ULTRAFAST MODULATION OF THE PULSE PROPERTIES

Let us now describe our approach to generate a pulse train with properties varying from pulse-to-pulse. We replace the sinusoidal phase modulation used in (1) by a two-tone temporal phase modulation:

$$\varphi'(t) = A_1 \cos(\omega_1 t) + A_2 \cos(\omega_2 t), \qquad (4)$$

where $A_1$ and $A_2$ are the amplitude of the sinusoidal phase modulation at angular frequencies $\omega_1$ and $\omega_2 = \omega_1 + \Delta\omega$, respectively. Combining these two sinusoidal modulations induces a beating characterized by fast oscillations with a frequency $\omega'_m$ (leading to a period $T_0'$) and by a slow envelope with a frequency $\omega_b$ (leading to a period $T_b$). Typical examples of these temporal beatings are plotted on Fig. 2(a). The resulting optical spectrum (Fig. 2b1) involves a set of spectral lines with typical spaces $\Delta\omega$ and $\omega_l$. The spectral shaping based on discrete $\pi/2$ phase shifts (Fig. 2b2) can efficiently be applied to each subset of spectral lines. As a consequence, we can expect that over the time scale of a fast temporal oscillation $T_0'$, the phase-to-intensity conversion will be qualitatively as efficient as the one obtained based on a purely sinusoidal modulation with an amplitude $A_m'$ provided by the envelope of the beating:

$$A_m'(t) = \sqrt{A_1^2 + A_2^2 + 2\,A_1\,A_2\cos(\Delta\omega\,t)}. \tag{5}$$

In order to get better understanding, it is interesting to discuss two limiting cases. When $A_1 \gg A_2$ (see for example panel a1 of Fig. 2), the envelope of the phase modulation can be approximated by a sinusoidal modulation (black mixed line):

$$A_m'(t) \simeq A_1 + A_2\cos(\Delta\omega\,t). \tag{6}$$

When $A_1 = 1.1$ rad, one can benefit from the linear approximation used in Fig. 1(b): the peak-power of the resulting pulses is impacted by the beating and will also follow a sinusoidal variation (see Fig. 3(a1), black mixed line), which amplitude is directly controlled by $A_2$. Pulse duration also changes, but for moderate values of $A_2$, the overall shape remains close to Gaussian pulses.

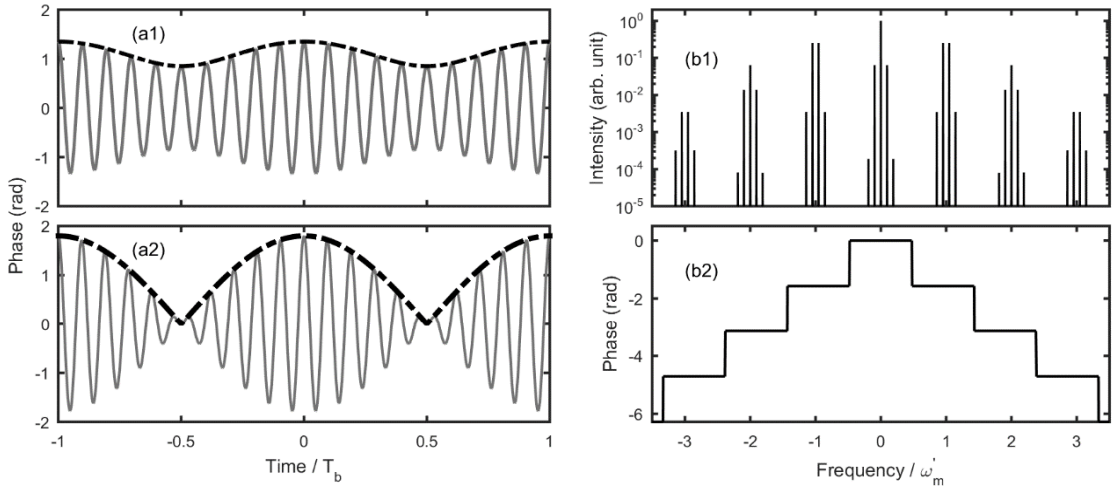

**Figure 2 (a)** Temporal phase profile resulting from the superposition of two sinusoidal phase modulation (grey solid line). Numerical results obtained for $A_1 = 1.1$ rad and $A_2 = 0.25$ rad (panel a1) or for $A_1 = .9$ rad and $A_2 = .9$ rad (panel a2). $\Delta\omega = 0.1\,\omega_l$. Envelope predicted by Eq. (6) and (8) are plotted with black mixed line. **(b1)** Optical spectrum obtained for $A_1 = A_2$. **(b2)** Spectral phase mask applied to the signal.

Another interesting case is when $A_1 = A_2$ (see Fig. 2a2). In this well-known case in wave physics [14, 15], Eq. (4) can be rewritten as :

$$\varphi'(t) = 2 A_1 \cos\left(\frac{\Delta\omega}{2} t\right) \cos\left(\frac{\omega_2 + \omega_1}{2} t\right), \quad (7)$$

leading to

$$A'_m(t) \simeq 2A_1 \left|\cos(\Delta\omega\, t)\right| \quad (8)$$

The fast oscillations have a frequency $\omega'_m = (\omega_1 + \omega_2)/2$. The envelope is modulated between 0 and $2 A_1$ at an angular frequency $\Delta\omega$. This results in rapidly varying pulses, with amplitude, duration, shape and extinction ratio experiencing major changes (see. Fig. 3b).

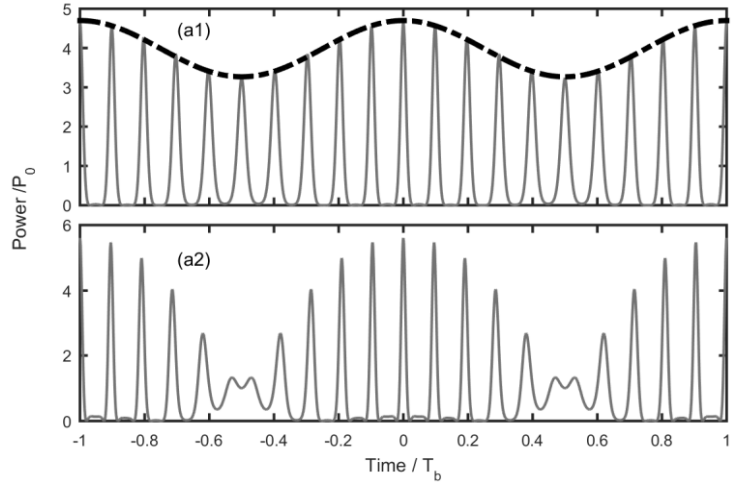

**Figure 3** Temporal intensity profiles resulting from the superposition of two temporal sinusoidal phase modulation after triangular spectral processing. Numerical results (grey line) are obtained at $\Delta\omega = .1\ \omega_1$ for $A_1 = 1.1$ rad and $A_2 = 0.25$ rad (panel a1) or for $A_1 = .9$ rad and $A_2 = .9$ rad (panel a2). Black mixed line represents the fluctuations of the peak-power based on Eq. (6) and the linear approximation from Fig. 1(b).

## 2.3 | PERFORMANCE ASSESSMENT

In order to more quantitively assess the ultrafast changes of the various properties experienced by the pulses, we have reported in Fig. 4, the temporal evolution of the peak-power, of the extinction ratio and of the duration. Results are evaluated over each period $T'_0$ (the $\Delta\omega$ being fixed here to 0.01 $\omega_1$) for different combinations of $A_1$ and $A_2$. The kurtosis excess (computed over the central part of the pulse, i.e. between two minima) also enables us to stress the changes in the temporal intensity profiles [16]. From Fig. 4, one can first note that for operation around $A_1$ fixed to 1.1 rad, the impact of the changes are directly linked to the amplitude of $A_2$. Whereas only moderate modulations are expected for $A_2 = 0.1$ rad, increasing $A_2$ to 0.5 rad significantly increases the range of fluctuations of the peak-power and durations, with peak-power that can vary by a factor of 2. The extinction ratio is also affected and some waveforms have an extinction ratio as poor as 10 dB. With fluctuations of the Kurtosis excess between -0.15 and 0.7, the temporal profile of each pulse can also be affected. When $A_1 = A_2 = 0.9$ rad, the range of the fluctuations is dramatically increased, with some waveforms presenting very low extinction ratios, indicating the presence of a strong continuous and coherent background for some pulses.

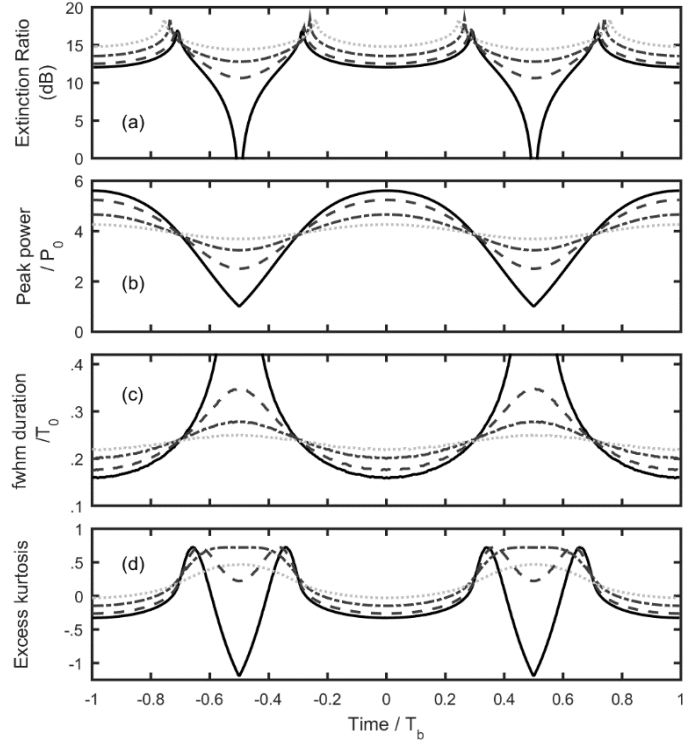

**Figure 4** Temporal evolution of different parameters for various combinations of parameters $(A_1, A_2)$ : (1.1, 0.1) (light grey dotted line) ; (1.1, 0.25) (grey mixed line)  ; (1.1, 0.5) (dark grey dashed line)  ; (0.9, 0.9) (black solid line). Evolution of the extinction ratio, of the peak-power, of the fwhm duration and of kurtosis excess (panels, a, b, c and d respectively). Results obtained for $\Delta\omega = 0.01 \; \omega_1$.

We summarized in Fig. 5 the evolution according to $A_2$ for $A_1 = 1.1$ rad of the amplitudes of fluctuation $\Delta P$ and $\Delta T$ of the peak-power and the fwhm duration respectively normalized to their average value. $\Delta P$ and $\Delta T$ are here defined as the difference between the maximal and minimal values that are recorded during the beating period $T_b$. We can note that both quantities evolve monotonously with $A_2$. For $A_2$ below 0.3 rad, the evolution obtained from numerical simulations follows closely a linear trend and can be predicted from the linear fits applied in panels (b) and (c) of Fig. 1. As a consequence, the level of jitters applied to the waveform is easily controlled.

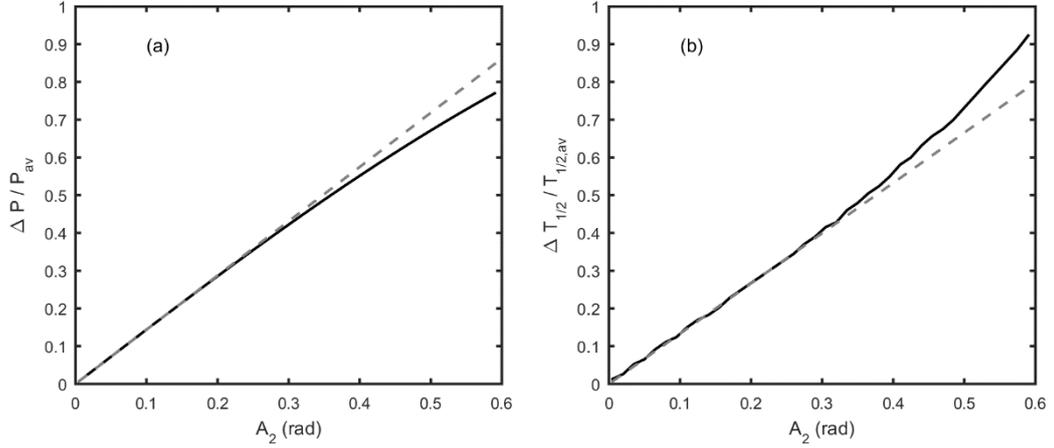

**Figure 5** Evolution of the amplitude jitter and the temporal width jitter (panel a and b respectively) according to the level of phase modulation $A_2$ for a fixed value of $A_1 = 1.1$ rad. Results of numerical simulations (solid black lines) are compared with a linear trend based on the linear fit applied in Fig. 1(b-c).

For $A_2 = 0.5$ rad, the range of fluctuations of the peak-power and the duration of the pulses can be as high as 67% of the average values. This leads to a major modification of the eye diagrams as stressed in Fig. 6. For moderate values of $A_2$ (panel a of Fig. 6), the eye remains clearly open, but with noticeable amplitude jitter. The probability distribution function of the peak-power exhibits a shape that is typical of sinusoidal fluctuations, as already measured for example in the past using jitter magnifiers [17]. For $A_2 = 0.5$ rad, both the amplitude jitter and the changes in the temporal duration contribute to the closure of the eye diagram. For a configuration where $A_2$ and $A_1$ are balanced, the level of fluctuations is greatly increased and the eye closed. The poor extinction ratio is also readily visible on the eye diagram.

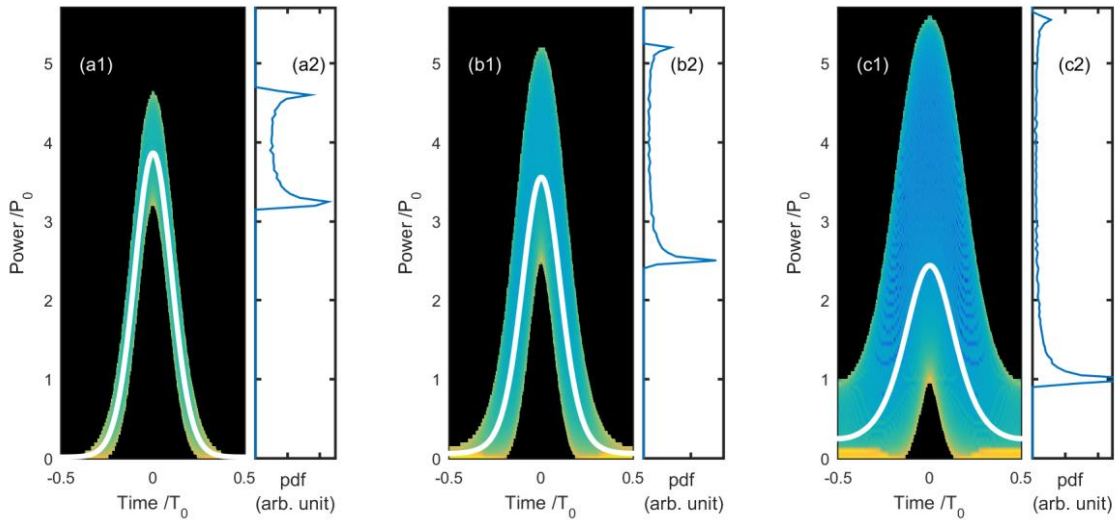

**Figure 6** Eye diagrams and probability distribution function of the peak-power of the pulse (panels 1 and 2 respectively) recorded for different combinations of phase parameters ($A_1$, $A_2$) : (1.1, 0.25) (panel a) ; (1.1, 0.5) (panel b) ; (0.9, 0.9) (panel c). Results obtained for $\Delta\omega = 0.01\ \omega_l$. The white line represents the waveform obtained from averaging.

## 3 | EXPERIMENTAL VALIDATION

### 3.1 | EXPERIMENTAL SETUP

The experimental setup is sketched on Fig. 3 and is based on devices that are commercially available and typical of the telecommunication industry. An external cavity laser delivers a continuous wave at 1550 nm that is then temporally phase modulated using two Lithium Niobate electro-optic devices driven by sinusoidal electrical signals at frequencies $\omega_1$ and $\omega_2$. The optical phase modulations have an amplitude $A_1$ and $A_2$. Note that a single phase modulator could be used if the electrical sinusoidal waveforms can be mixed in the electrical domain. For these proof-of-principle experiments, $\omega_1$ has been fixed to 10 GHz. A linear spectral shaper (Finisar Waveshaper) based on liquid crystal on silicon technology [18] is then used to apply the discrete spectral phase shifts of π/2 between two successive components. In order to ensure enhanced environmental stability, polarization-maintaining components have been used. Moreover, since the principle of operation is purely linear, no Erbium doped fiber amplifier is required, thus limiting the source of detrimental noise. The reshaping process is here quite energy efficient, since the optical losses are restricted to the insertion losses of the phase modulators and of the spectral shaper.

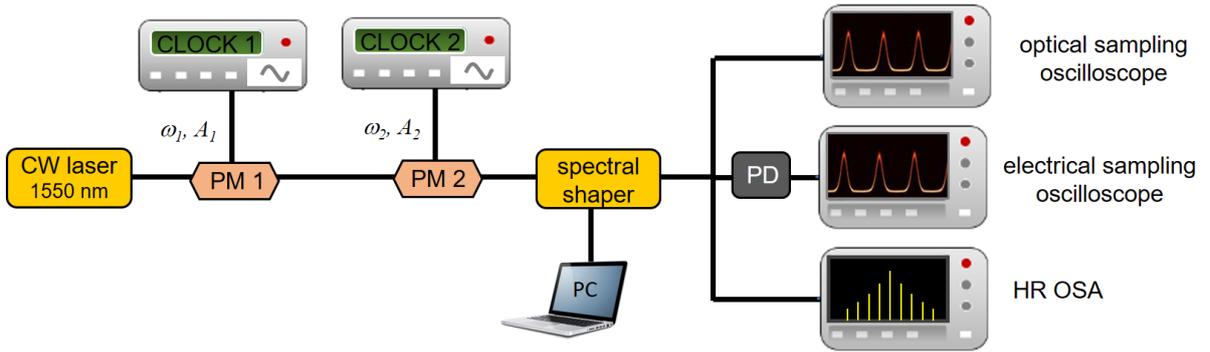

**Figure 7** Experimental setup. CW : continuous wave ; PM : phase modulator ; HR OSA : high resolution optical spectrum analyser ; PD : Photodiode.

The resulting signal is directly recorded by means of a high-speed optical sampling oscilloscope (1 ps resolution), an electrical sampling oscilloscope (50 GHz bandwidth) and a high-resolution optical spectrum analyser (5 MHz resolution).

### 3.2 | EXPERIMENTAL RESULTS

In order to validate our concept, we have first recorded the temporal waveforms that are generated when temporal phase modulations have frequencies of $\omega_1 = $ 10 GHz and $\omega_2 = $ 11 GHz, associated with an amplitude of modulation of $A_1 = $ 1.1 rad and $A_2 = $ 0.25 rad (panel a1) or $A_1 = A_2 = $ 0.9 rad (panel a2). Results are recorded on an optical sampling oscilloscope and plotted on Fig. 8(a). The experimental waveforms outline that modification of the temporal pulse properties on a pulse-to-pulse basis is efficiently achieved: in both cases, each pulse of the 10 pulses sequence exhibits features that are different from its nearest neighbor. Changes in the peak-power are associated with changes of the temporal duration and the residual background can also vary. The excellent agreement that is obtained with the numerical simulations (see dotted grey line) stresses the accurate control that can be achieved by simply adjusting the amplitude of the phase modulation. The typical optical spectrum as shown in panel (b) of Fig. 8 confirms the high level of the optical signal to noise ratio, confirming that the observed fluctuations are fully deterministic and cannot be linked to any amplified spontaneous emission.

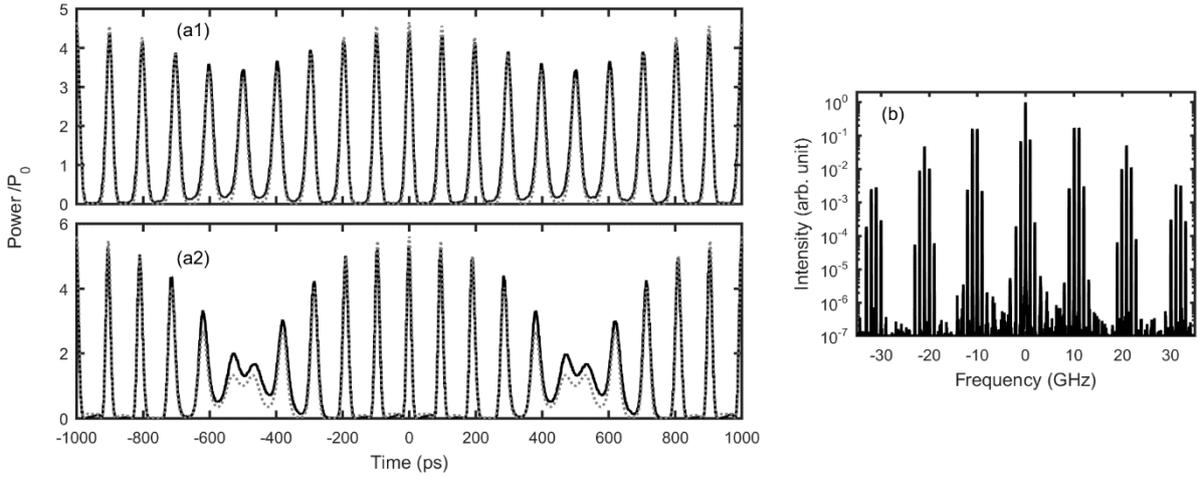

**Figure 8** (a) Temporal intensity profiles resulting from the superposition of two sinusoidal phase modulation and spectral processing. Results obtained for $A_1 = 1.1$ rad and $A_2 = 0.25$ rad (panel a1) or for $A_1 = .9$ rad and $A_2 = .9$ rad (panel a2). Experimental results recorded on the optical sampling oscilloscope (solid black line) are compared with numerical simulations (grey dotted lines) (b) Optical spectrum recorded for $A_1 = .9$ rad and $A_2 = .9$ rad. All results are recorded for $\omega_1 = 10$ GHz and $\omega_2 = 11$ GHz.

We performed additional measurement using a longer sequence of modulated pulses (i.e. with $\omega_1 = 10$ GHz and $\omega_2 = 10.1$ GHz) for a fixed value of $A_1 = 1.1$ rad and values of $A_2$ spanning from 0.1 rad to 0.6 rad. Benefiting from the electrical sampling oscilloscope, we were able to record the resulting 10 ns pulse train. Summary of the fluctuation levels of both the peak-power and the temporal duration is presented in Fig. 9. Both types of fluctuations increase with the amplitude of $A_2$. The experimental trends qualitatively reproduce the numerical results described in Fig. 5. The deviation that is observed can be mainly ascribed to the finite bandwidth of our optoelectronic detection.

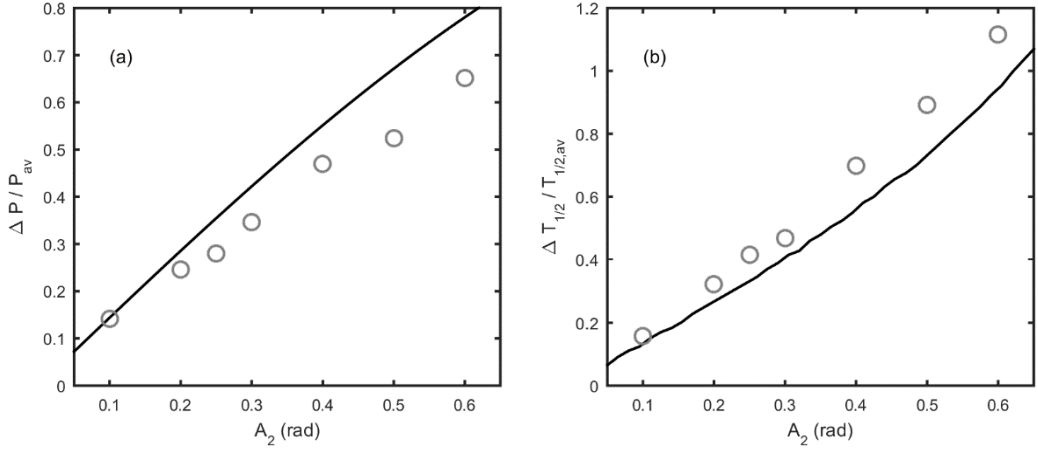

**Figure 9** Evolution of the amplitude jitter and the temporal width jitter (panel a and b respectively) according to the level of phase modulation $A_2$ for a fixed value of $A_1 = 1.1$ rad. Results of numerical simulations (solid black lines) are compared with the experimental results measured on the digital sampling oscilloscope for $\omega_1 = 10$ GHz and $\Delta\omega = 0.1$ GHz.

As a final measurement, we reconstructed eye diagrams from the experimental measurements. Results are provided on Fig. 10 for three different levels of jitters, corresponding to $A_2 = 0.25$ rad or 0.5 rad with $A_1$ being fixed to 1.1 rad (panel a and b respectively), or to $A_2 = A_1 = 0.9$ rad (panel c). Once again, the experimental results are found in qualitative agreement with the numerical simulations presented Fig. 6. They therefore confirm the various predictions regarding the different regimes that can be observed, ranging from slight fluctuations of the temporal width and peak-power to strong distortions of the temporal properties of the generated waveform.

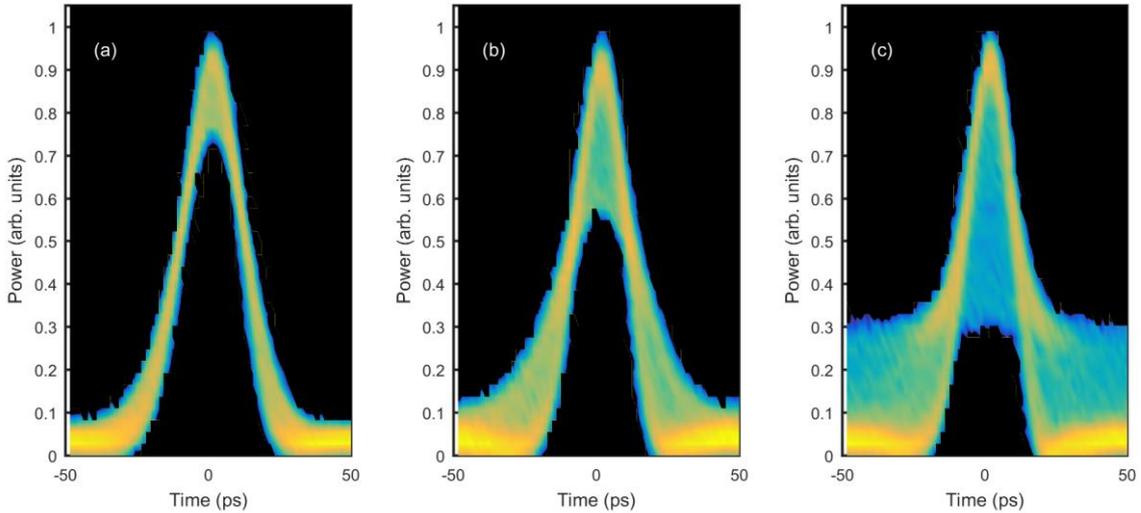

**Figure 10** Experimental eye-diagrams recorded for different combinations of phase parameters $(A_1, A_2)$ : (1.1, 0.25) (panel a) ; (1.1, 0.5) (panel b) ; (0.9, 0.9) (panel c). Results obtained for $\Delta\omega = 0.01\ \omega_1$.

## 4 | CONCLUSIONS

In order to conclude, we have theoretically proposed and numerically validated an all-optical scheme to generate optical pulse trains with peak-powers and durations that may significantly vary from pulse-to-pulse. Our approach relies on a shaping of the spectral phase with $\pi/2$ phase shifts. This enables an efficient phase-to-intensity conversion of a temporal phase modulation based on a two-tone sinusoidal beating. Experiments fully confirm the ability of our approach to efficiently generate a train made of pulses with properties that vary from pulse-to-pulse. The levels of jitters can be accurately controlled and significant levels of fluctuations can be recorded.

Our proof-of-principle experiments have been carried out at a repetition rate of 10 GHz. However, as the principle of a stable pulse train generation has also been validated at 40 GHz [8], we can expect that similar results and jitters could be achieved at 40 GHz. As suggested in [19], our approach should also be suitable for multi-wavelengths operation. In the present work, we have involved two purely sinusoidal phase modulations. By taking advantage of one frequency-swept sinusoidal wave, it could be possible to introduce a controlled fluctuation of the temporal position of the pulse.

## ACKNOWLEDGEMENTS

We acknowledge the support of the Institut Universitaire de France (IUF), the Bourgogne-Franche Comté Region, the French Investissements d'Avenir program and the Agence Nationale de la Recherche (ANR-11-LABX-01-01). We thank Julien Fatome and Bertrand Kibler for fruitful discussions or technical help, as well as Kamal Hammani for initial comments. The article has benefited from the PICASSO experimental platform of the University of Burgundy.

## CONFLICT OF INTEREST

The author declares no potential conflict of interest.